\documentclass[manuscript,acmsmall]{acmart}

\def\BibTeX{{\rm B\kern-.05em{\sc i\kern-.025em b}\kern-.08emT\kern-.1667em\lower.7ex\hbox{E}\kern-.125emX}}

\setcopyright{acmcopyright}
\copyrightyear{2020}
\acmYear{2020}
\acmConference[ASSET '20]{ASSET '20: The 22nd International ACM SIGACCESS Conference on Computers and Accessibility, October 26--28, 2020, Athens, Greece (Virtual Conference: Poster)}
\acmPrice{15.00}
\acmDOI{10.1145/1122445.1122456}
\acmISBN{978-1-4503-9999-9/18/06}

\begin{document}

\title{Substituting Restorative Benefits of Being Outdoors through Interactive Augmented Spatial Soundscapes}

\author{Swapna Joshi}
\affiliation{%
  \institution{Indiana University Bloomington}}
\email{swapna@iu.edu}

\author{Kostas Stavrianakis}
\affiliation{%
  \institution{Indiana University Bloomington}}
\email{kstavria@iu.edu}

\author{Sanchari Das}
\affiliation{%
  \institution{Indiana University Bloomington, University of Denver}}
\email{sancdas@iu.edu}

\renewcommand{\shortauthors}{Joshi and Stavrianakis et al.}

\begin{abstract}
Geriatric depression is a common mental health condition affecting majority of older adults in the US. As per Attention Restoration Theory (ART), participation in outdoor activities is known to reduce depression and provide restorative benefits. However, many older adults, who suffer from depression, especially those who receive care in organizational settings, have less access to sensory experiences of the outdoor natural environment. This is often due to their physical or cognitive limitations and from lack of organizational resources to support outdoor activities. To address this, we plan to study how technology can bring the restorative benefits of outdoors to the indoor environments through augmented spatial natural soundscapes. Thus, we propose an interview and observation-based study at an assisted living facility to evaluate how augmented soundscapes substitute for outdoor restorative, social, and experiential benefits. We aim to integrate these findings into a minimally intrusive and intuitive design of an interactive augmented soundscape, for indoor organizational care settings. 
\end{abstract}




\begin{teaserfigure}
 \includegraphics[height=4.5cm, width=\textwidth]{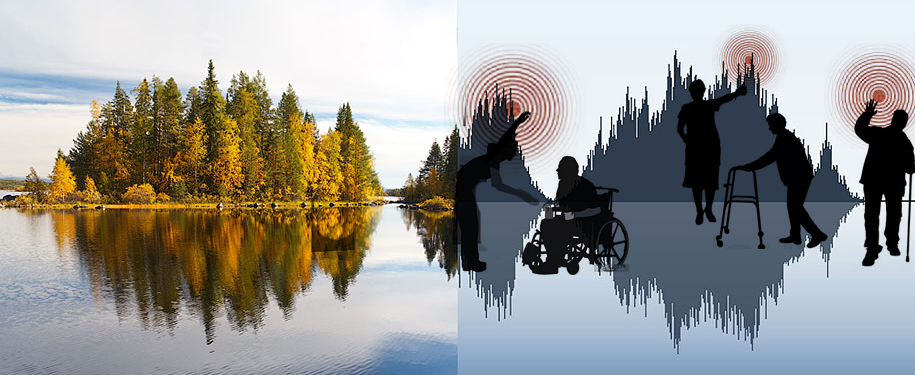}
 \label{fig:teaser}
\end{teaserfigure}

\maketitle
\pagestyle{plain}
\section{Introduction}
Nearly one in five older adults in the US currently experience some form of mental illness, of which 15-20\% are affected by depression~\cite{blazer2012mental,ciechanowski2004community}. Grounded in the Attention Restoration Theory (ART)~\cite{ohly2016attention}, prior research suggests that exposure to restorative settings such as the natural environment, can treat and manage depression by reducing mental fatigue and stress~\cite{kaplan1989experience,pearson2014great}. However, researchers have identified how older adults' mobility and sensory acuity decline over time, showing deficits in the coordination of bi-manual and multi-joint movements limiting their outdoor activities~\cite{kono2004frequency}. Also, many older people receive care and treatment in assisted living and organizational care settings, which provide limited resources and support for outdoor activities~\cite{jang2006depressive}. For this reason, older people have less access to the pleasure and enjoyment of sensory engagement with nature and its associated restorative benefits on well-being and quality of life~\cite{orr2016older}. 

To enable older adults to overcome these barriers, including mobility and sensory limitations and organizational care resource shortage, there is a growing interest in the development of persuasive mobile and wearable technology~\cite{cozza2017ubiquitous} for supporting outdoor activities. However, new-age techniques are yet to provide adequate solutions that could address the deficit of older people's everyday sensory exposure to nature. Thus, our research aims to explore the restorative benefits of indirect exposure to nature using an augmented spatial soundscape~\cite{hong2017spatial} experience for older adults with depression in care facilities. By applying the principles of ART, we aim to design an intuitive experience for older adults to benefit from the restorative and social benefits of outdoors through an interactive spatial augmented soundscape.

\section{Background and Motivation}
\subsection{Geriatric Depression and ART}
Older adults in an assisted living encounter multiple life transitions and may be vulnerable to depressive illness. Walson et al. studied a geographically diverse sample of assisted-living residents, where it indicated that depression is common among residents of organizational care facilities~\cite{watson2003depression}. 

Among other factors associated with the improvement of mental health, the positive contribution of nature exposure has been widely studied both from a physical and mental health perspective~\cite{berman2008cognitive,berman2012interacting}.
The Attention Restoration Theory is possibly the most common framework with behavioral studies looking at humans' well-being and exposure to nature. 
According to ART, for cognitive restoration to be achieved, individuals should engage in involuntary attention through exposure to restorative environments, such as nature. For an environment to be restorative, it should be a) extent, b) being away, c) offers fascination, and d) be compatible with the user~\cite{kaplan1995restorative}.

Being outdoors and being active have also been related to fewer depressive symptoms for elderly living in assisted living and care facilities~\cite{kerr2012relationship}. However, a study of outdoor spaces in nursing homes suggested significant variation in outdoor amenities and access to those amenities across facilities, with the majority of the nursing homes having no outdoor amenities~\cite{cutler2006great}. Further, a study revealed that although some facilities had the necessary infrastructure to support outdoor activities, the physical limitations of the elderly acted as a barrier from going outdoors. Even the population that was physically able to go outdoors chose to do so less than once a month~\cite{van2020understanding}. 

Further, a review of interventions for depression has called for more research to explore simple, portable, and cost-effective interventions that go beyond expensive conventional treatments such as psychotherapy or medication, emphasizing the role of technology to reach a large population in need of services~\cite{kazdin2011rebooting}. 

\subsection{ART and Technology}
Despite the known benefits of contact with the natural outdoor environment, elderly people living in care institutions have less opportunity to avail of such restorative benefits for their depression condition. 

Technology has been used to deliver the visual aspects of the outdoors, such as digital images of nature for cognitive restoration~\cite{emfield2014evaluating,valtchanov2010restorative}. However, it appears that in addition to the visual elements of interacting with nature and its restorative and therapeutic benefits, calming nature sounds and even outdoor silence can have similar restorative effects~\cite{krzywicka2017restorative,jahncke2015effects}. This aligns with the different tenets of ART especially, the two tenets of being away and fascination~\cite{kaplan1995restorative}, for example, some environmental sounds can increase the feeling of being away in another environment (e.g., sounds from a forest) and increase fascination through content variations (e.g., the twitter of birds, rippling water). 

Considering the above, individuals with depression could benefit from the experience of nature sounds in their care facilities. Modern advancements in immersive technology would allow for the complete customization and creation of natural restorative environments~\cite{baus2014moving}. However, much remains unexplored in the areas of immersive and interactive virtual soundscapes, especially in the context of depression. Augmented reality spatial soundscapes could allow for immersive experiences while requiring minimally invasive technologies~\cite{duezguen2020towards}, such as with a system of bone conduction headphones~\cite{buroojy2009bone}, and gesture-based controls~\cite{gummeson2014energy}. 

Older adults are found to struggle with technologies that are not designed to accommodate their needs or adapted to their specific cognitive, sensory, and motor abilities~\cite{stossel2009familiarity,das2019towards}. Studies have shown how the physical presence of the heavy technological equipment~\cite{das2020don}, in combination with the psychological and mental fear of the isolating experiences they could bring, leads to high levels of technology rejection in elderly users~\cite{coleman2016gesture}. However, technologies such as Bone-conduction headphones and gesture control technologies for elderly users have shown a high acceptance rate, and proper use with the requirement of little familiarity~\cite{stossel2009familiarity,chen2013gesture}. Bone conduction Headphones have speaker pads that rest on the temples of the user and allow for picking up some sound from the surrounding contexts, thus layering the real-world environment with virtual audio, rather than isolating the user from the real world and social surroundings. Gesture control technologies, on the other hand, strengthen the immersive experience and allow for unique intuitive ways to interact with the technology~\cite{coleman2016gesture}. Thus, by mapping hand gestures, the technology enables preserving user autonomy by letting the users mediate and customize the interface in real-time. 

Below, we present our research design that uses 1) strong underlying theoretical principles of ART, 2) potential of intelligent spatialized augmented soundscapes, and 3) the interaction and control technologies like gesture-based ring controls and bone conduction headsets, to provide older adults with the restorative and social benefits of outdoors within their indoor environments.

\section{Research Proposal}
Through semi-structured interviews of 8-10 older adults in an assisted living center, first, we aim to learn their preferences for natural sounds and understand how they associate different sounds with memories, experiences, or moods while they are in an indoor social environment of assisted living. This would allow us to learn how different sounds could provide them with restorative benefits discussed in ART and how they could engage in the social experience with others using these soundscapes, through their shared preferences and experiences.
We then plan to create a database of natural sounds based on these interviews to develop four to six spatial soundscapes, comparable to the sensory experience of natural outdoor soundscapes. We then plan to test these spatial soundscapes with our participants using bone conduction headphones. As they are listening to these soundscapes, we plan to ask them a few questions based on ART theory to understand if they perceive the spatial soundscapes as restorative and enjoyable. 

Further, we would ask our participant older adults to suggest their intuitive ways of interacting with different sounds through gestures, if they had to, such as to repeat listening to a sound, stop listening to a sound or bring a sound closer. For example, we will ask them questions like- ''Do you hear that bird chirping? (Participant responds yes/no) Where do you think the sound is coming from? (Participant shows direction) How far do you think that bird is (Participant answers if they think the bird is close-by or far away) How would you bring that bird closer using gesture? OR Could you imagine using your hands to bring that bird closer? How would you shoo away the bird to stop listening to its sound? (Participant responds by gesture).'' These in-situ responses of older adults will be recorded as observations. Using these observations and interviews, we hope to build our Wizard of Oz prototype of gesture-based interaction through a gesture control ring for older adults to intuitively interact with the natural spatial soundscape and share the interaction space.

The prototype will be tested in an indoor social space of an assisted living facility, where three-four participating older adults will wear a prototype ring that they would be told as recording their gestures. They will all listen to a shared spatial natural soundscape and interact with it intuitively using gestures. Their shared interactions will be remotely executed by Wizard of Oz researchers and video recorded as observations. Finally, we will evaluate our WoZ prototype and its restorative and social benefits, using in-situ interviews of older adults and observing their gesture-based interactions to understand challenges and limitations for further iterations to our prototype. 

The spatial audio interface's unique design would allow for communal experiences as opposed to most mixed reality solutions that are individual-oriented. The ring-based real-time gesture interaction with the augmented natural soundscapes would allow for engaging and fascinating experiences for others who share the interaction space and may lead to social conversations opportunities. For example, while listening to the augmented spatial soundscape, with sounds of birds, stream or the rustling of the leaves from different directions, an interactor could use intuitive gesture in the direction of the sound, such as by moving their arms to throw a stone in the water to shape the experience and share it with others navigating the soundscapes. 
\begin{figure}
    \centering
    \includegraphics[height=3.8cm,width = \textwidth]{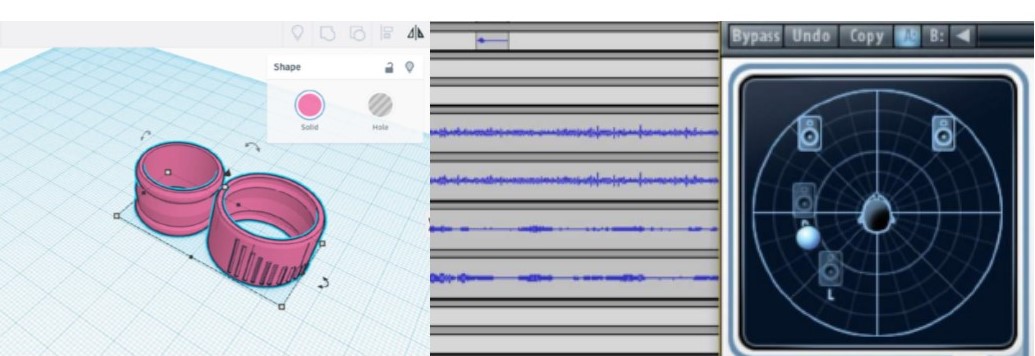}
    \caption{Ongoing prototype of the gesture ring and 3D soundscapes.}
    \label{fig:1}
    \vspace{-5mm}
\end{figure}

\section{Expected Contribution}
Our study would be the first-ever study of augmented interactive soundscapes for promoting restorative mental health for elderly people with depression in organizational care settings. 
Unlike available solutions that require users to learn and get acquainted with the technology~\cite{baus2014moving}, our proposed interactive spatial soundscape, provides an intuitive and minimally intrusive way to connect with nature (using headphones and ring-based gesture controllers), while allowing connection with the social surroundings, rather than disconnecting or isolating the user. As an alternative intervention for depression, our technology design could prove feasible and implementable as it would be low cost, especially for our target shared and communal care organization settings. 

\section{Broader Significance}
Recently many assisted living and long-term care facilities for older adults in the US became hotspots for COVID-19, exposing frail older people with multiple comorbidities and functional limitations to the highest risk. As the virus swept through the nation, these facilities locked down. They instituted tight restrictions on activities that might expose older adults to others, including outdoor activities and the sensory benefits. Such canceling of any possible outdoor activities have made them older adults unfairly constrained. With the possibility of a second wave and seasonality of COVID-19, the changes and challenges to outdoor activities in long-term care facilities work may remain and even heighten after the lockdown and distancing restrictions start easing out. Such challenges and risks are not limited to older adults. They apply to others who are at high risk, experiencing anxiety and stress, or have disabilities that require them to stay indoors in the near future. Now more than ever, bringing sensory and social experiences of outdoors to indoors seems beneficial at a broader level. Further, the applicability of the study, broadly in interactive spatial audio-based navigation, cannot be undermined to those who are visually impaired and unable to benefit from visual AR/VR technologies. 

\section{Acknowledgement}
We want to thank Indiana University's Big Idea - Prevention Insights for their initial feedback on this project; Secure and Privacy Research in New-Age Technology (SPRINT) Lab, University of Denver; and Human and Technical Security (HATS) Lab, Indiana University. Any opinions, findings, and conclusions or recommendations expressed in this material are those of the author(s). They do not necessarily reflect the views of the University of Denver, or Indiana University

\bibliographystyle{ACM-Reference-Format}
\bibliography{ARXIV_ASSETS_AR}

\end{document}